%% file: sample-authordraft.tex
\documentclass[sigconf]{acmart}

\AtBeginDocument{%
  \providecommand\BibTeX{{%
    \normalfont B\kern-0.5em{\scshape i\kern-0.25em b}\kern-0.8em\TeX}}}
\setcopyright{none}

\usepackage[most]{tcolorbox}
\usepackage{multirow}
\usepackage{titlesec}
\usepackage{enumitem}
\setlist{nosep} %

\titlespacing*{\paragraph}
{8pt}{0\baselineskip}{0.5\baselineskip}

\tcbset{textmarker/.style={%
        enhanced,
    parbox=false,boxrule=0mm,boxsep=0mm,arc=0mm,
        outer arc=0mm,left=1mm,right=1mm,top=7pt,bottom=7pt,
        toptitle=1mm,bottomtitle=1mm,oversize}}

\newtcolorbox{importantBox}{textmarker,
    borderline west={2pt}{0pt}{red},
    colback=red!10!white}

\newcommand{\important}[1]{\begin{importantBox} \textbf{} #1 \end{importantBox}}

\newcommand{\tsc}[1]{\textsuperscript{#1}} %

\begin{document}
\newcommand{\todo}[1]{\textbf{\textcolor{red}{ToDO: #1}}}
\newcommand{\wei}[1]{\textcolor{orange}{wei: #1}}
\newcommand{\lin}[1]{\textcolor{blue}{#1}}
\newcommand{\fixme}[1]{\textcolor{red}{#1}}
\newcommand{\cameraready}[1]{\textcolor{black}{#1}}
\newcommand{\recheck}[1]{\textcolor{green}{ #1}}
\newcommand{\revision}[1]{\textcolor{black}{ #1}}
\newcommand{\tb}[1]{\textcolor{red}{bissyande$\blacktriangleright$#1$\blacktriangleleft$}}
\newcommand{\jk}[1]{\textcolor{red}{Jacques $\blacktriangleright$#1$\blacktriangleleft$}}

\newcommand{\answer}[2]{
  \begin{tcolorbox}[enhanced, left=3mm,right=3mm,
    colback=gray!10, colframe=gray!80, boxrule=0pt,
    borderline west={4pt}{0pt}{gray!90},
    ]
    \textbf{Answer for RQ#1:}
    #2
    \end{tcolorbox}
}

\def\figref#1{Figure~\ref{fig:#1}}
\def\figlabel#1{\label{fig:#1}\label{p:#1}}
\def\tabref#1{Table~\ref{tab:#1}}
\def\tabsref#1{Tables~\ref{tab:#1}}
\def\tablabel#1{\label{tab:#1}\label{p:#1}}
\def\Secref#1{\S\ref{sec:#1}}
\def\secref#1{\S\ref{sec:#1}}
\def\seclabel#1{\label{sec:#1}}
\def\qref#1{Eq.~\ref{eqn:#1}}
\def\eqrefn#1{\ref{eqn:#1}}
\def\eqsref#1#2{Eqs.~\ref{eqn:#1}/\ref{eqn:#2}}
\def\eqlabel#1{\label{eqn:#1}}
\def\subsp#1{P_{\mbox{{\scriptsize\rm #1}}}}
\def\appref#1{Appendix~\ref{sec:#1}}

\title{Open-Source AI-based SE Tools: Opportunities and Challenges of Collaborative Software Learning}

\author{Zhihao Lin\tsc{1}, Wei Ma\footnote{}\tsc{2}, Tao Lin\tsc{3}, Yaowen Zheng\tsc{2}, Jingquan Ge\tsc{2}, Jun Wang\tsc{4}, Jacques Klein\tsc{4}, Tegawende Bissyande\tsc{4}, Yang Liu\tsc{2},Li Li\tsc{1}}\thanks{*contributed equally to this work.}
\affiliation{
	\institution{1. Beihang University, China}
	\institution{2. Nanyang Technological University, Singapore}
	\institution{3. Westlake University, China}
	\institution{4. University of Luxembourg, Luxembourg}
	\country{}
}

\begin{abstract}

Large Language Models (LLMs) have become instrumental in advancing software engineering (SE) tasks, showcasing their efficacy in code understanding and beyond. Like traditional SE tools, open-source collaboration is key in realising the excellent products. However,  with AI models, the essential need is in data.
The collaboration of these AI-based SE models hinges on maximising the sources of high-quality data. However, data especially of high quality, often holds commercial or sensitive value, making it less accessible for open-source AI-based SE projects. This reality presents a significant barrier to the development and enhancement of AI-based SE tools within the software engineering community. Therefore, researchers need to find solutions for enabling open-source AI-based SE models to tap into resources by different organisations. 
Addressing this challenge, our position paper investigates one solution to facilitate access to diverse organizational resources for open-source AI models, ensuring privacy and commercial sensitivities are respected. We introduce a governance framework centered on federated learning (FL), designed to foster the joint development and maintenance of open-source AI code models while safeguarding data privacy and security. Additionally, we present guidelines for developers on AI-based SE tool collaboration, covering data requirements, model architecture, updating strategies, and version control. Given the significant influence of data characteristics on FL, our research examines the effect of code data heterogeneity on FL performance. 
We consider 6 different scenarios of data distributions and include 4 code models. We also include 4 most common federated learning algorithms. \revision{Our experimental findings highlight the potential for employing Federated Learning in the collaborative development and maintenance of AI-based software engineering models.} We also discuss the key issues to be addressed in the co-construction process and future research directions.

\end{abstract}


\maketitle
\renewcommand{\shortauthors}{Z. Lin and W. Ma et al.}

\section{Introduction}
\label{sec:introduction}

\input{sections/introduction}

\section{Background}
\label{sec:background}
\input{sections/background_and_related_work}

\section{Proposal}
\label{sec:motivation}
\input{sections/motivation}

\section{Experiment Design}
\label{sec:expDesign}

\input{sections/experiment}

\section{Results}
We first demonstrate the model performance under federated learning using different data distributions. In the end, we answer our research questions~(\secref{answer}) based on the experiment results.
\label{sec:results}
\input{sections/results}

\section{Challenges and Opportunities}
\label{sec:challenges}
\input{sections/challenges}

\section{Conclusion}
\label{sec:conclusion}
In this study, we delve into an emerging framework for the collaborative development of open-source code models, the code model-sharing mechanism based on federated learning. This framework aims to serve as a new paradigm for managing open-source AI-based SE models and to foster the joint construction and refinement of code models while ensuring data privacy protection. Through this paradigm, we have detailed the specific impact of code data heterogeneity on model performance and identified key issues and challenges. Moreover, we discuss some potential challenges and solutions that might be faced. For example, we should intensify research on implementing more secure and reliable comprehensive encryption methods tailored to code characteristics.
Regarding the contribution incentive mechanism for open-source models, we propose using a token-based contribution method to encourage more developers and participants to join this collaborative effort. At the same time, we also highlight the challenges of managing intellectual property rights for models in the implementation process. We look forward to further research and discussion in this field.

\bibliographystyle{ACM-Reference-Format}
\bibliography{sample-base}
\end{document}

%% file: sections/introduction.tex
Large language models~(LLM) are profoundly transforming various areas of software engineering~(SE).
LLMs have significantly influenced the methodologies of SE systems~\cite{fan2023large}. Taking MetaGPT~\cite{hong2023metagpt} as an example, this tool can generate code based on human language descriptions. It simulates the basic process of software development by mimicking the collaboration among team members and integrating the programming capabilities of LLM. Although the development mode of MetaGPT is relatively simple compared to agile development~\cite{abrahamsson2017agile} and lacks measures for security and quality assurance, it has demonstrated the potential of LLMs in the field of software engineering. 
Researchers have proposed many code models, which can be grouped into two categories according to the model size: pre-trained code models~(small) and foundational code models~(large). Pre-trained code model series (like CodeBERT~\cite{feng2020codebert}, CodeGPT~\cite{chen2021evaluating}, and GraphCodeBERT~\cite{guo2020graphcodebert}) and foundational code models (like CodeLlama~\cite{roziere2023code}, StarCoder~\cite{li2023starcoder}, and CodeT5+~\cite{wang2023codet5+}) have attracted much attention. Researchers have trained expert models based on pre-trained models or foundational code models, which are also emerging, like code repair models~\cite{jiang2021cure}. These code models have gradually become basic tools, called AI-based SE tools, in the field of software engineering~\cite{xu2022systematic} such as the bug fixing tool SWE-agent~\cite{yang2024sweagent}.

The current open-source code model is mainly developed and published by a single team based on open-source data. However, three significant limitations exist in how open-source models are developed and shared: \textit{limited access to high-quality code data}, %
\textit{lacking community strong support} and \textit{training hardware resources}.

\textit{\textbf{First}}, it is well known that data quality is crucial for the performance of AI models~\cite{ntoutsi2020bias}. Currently, most open-source code models rely on publicly available code datasets. However, the growth rate of high-quality open-source code data has not kept up with the pace at which LLM capabilities are improving. According to the law of scaling~\cite{kaplan2020scaling}, the performance of a model is directly proportional to the amount of high-quality data. 
Therefore, to develop more powerful code models, there is an urgent need for more high-quality code data.

\textit{\textbf{Second}}, despite their widespread application in many areas of software engineering, such as vulnerability detection~\cite{li2018vuldeepecker}, they still lack the strong open-source community support typical of traditional software engineering tools.
These open-source models also resemble isolated information islands, where individual entities independently complete the training and release of models using open-source data. In an era when large language models (LLMs) are beginning to reshape the field of software engineering, this pattern needs to be updated.  \revision{For example, when an individual notices that current vulnerability detection systems underperform with certain vulnerabilities, they retrain the model using additional data, addressing the issue. However, this enhanced model is only available in the new location and has not been assessed for its ability to maintain its original performance levels. Similarly, another individual implemented comparable upgrades for different flaws in the model. Consequently, users need help finding these updated models and may view these unverified improvements with skepticism.}

\textit{\textbf{Third}}, the current unipartite participation model leads to low economic efficiency, as different developers or teams conduct redundant training on the same dataset, and the training process of AI models consumes a significant amount of resources, e.g., electricity~\cite{chen2021evaluating}.

Despite the daunting challenges faced in intelligent software engineering, we can still explore solutions. For the first challenge, commercial IT companies or organizations hold high-quality code data and are reluctant to share it, severely limiting industry innovation and development. Based on our collaborative experience with enterprises, IT companies have a strong sense of protection for even a single line of code commentary, let alone the complete code. If a mechanism could be designed to encourage these companies to share their valuable code assets, the development of intelligent software engineering would be significantly propelled. Furthermore, for the second and third challenges, establishing a model-building process that involves multiple parties and integrates different datasets is more efficient and environmentally friendly than training models in isolation. 

Therefore, we focus on federated learning (FL)~\cite{konecny2016federated} — a distributed machine learning framework aimed at enabling multi-party collaboration under the auspices of data privacy protection. Federated learning safeguards data privacy and compliance, and significantly enhances AI model performance through collaborative modeling. There are two related works. FedCSD~\cite{alawadi2023fedcsd} uses federated learning for detecting code smells while preserving data privacy.
ALMITY~\cite{10.1109/TSE.2023.3347898}, designed to address the challenges of applying academic models to real-world industrial applications. Both do not discuss how code data distribution affects federated learning and do not consider the potential ability of federated learning to change the community of open-source code models. Our work has deeply researched the impact of different data distributions on federated learning and outlines how we can use federated learning to govern open-source code models. We consider the different code data distribution and include several pre-trained code models and one large foundation model in our experiments. \revision{We include 4 most common FL strategies for the model aggregation.}

In this work, we outline the co-constructed code-model framework and we explore how data distribution among different participants affect the code model performance in this cooperative framework. 
\revision{We demonstrate through experiments that federated learning can achieve performance similar to the centralized training. We also find that FL can achieve better results than individual participants training on their own data while ensuring data privacy for each participant.}
The experimental code is available~\cite{codelink}. 

In summary, our study makes the following contributions:
\begin{enumerate}
    \item In view of the current problems of code data sharing and code model maintenance, we discussed a framework for collaborative AI-based SE tools building based on federated learning.
    \item  We undertake a thorough investigation to assess the influence of data heterogeneity on the outcomes of federated learning models. This comprehensive study aims to understand how  heterogeneity in data can affect the model performance and generalization. %
    \item   Our experimental results strongly supports the potential use of federated learning in bringing together various companies to collaborate on the development of intelligent software engineering, thereby promoting the advancement of this field.
\end{enumerate} 

\paragraph{\textbf{Roadmap}} The paper is organized by the following way. \secref{background} introduces the background. \secref{motivation} demonstrates our motivation and our solution. \secref{expDesign} introduces the data distribution strategies, the dataset, the models and the FL algorithms we used. \secref{results} shows our experimental results. \secref{challenges} state the challenges and opportunities. \secref{conclusion} concludes our work.

%% file: sections/background_and_related_work.tex
\subsection{Large Language Models in SE}

The pre-trained models have significantly enhanced task performance in natural language processing, attributed to their excellent generalization capabilities~\cite{kim2022broad}. Researchers have adapted pre-trained transformer models to code data in the software engineering domain~\cite{von2022validity}. Based on their pre-training strategies and architectural designs, these models are categorized into three types: autoencoding, autoregressive, and sequence-to-sequence (Seq2Seq) models.
Autoencoding models leverage the transformer encoder and adopt strategies like masked language modelling (MLM)~\cite{sinha2021masked}, which hides certain parts of the code and uses the surrounding context to predict them. This allows the utilization of future token information for current predictions. For instance, CodeBERT~\cite{feng2020codebert} is trained on the CodeSearchNet~\cite{husain2019codesearchnet} dataset and features unique data flow input in addition to regular code and natural language inputs, as demonstrated by GraphCodeBERT~\cite{guo2020graphcodebert}.
Conversely, autoregressive models, exemplified by CodeGPT~\cite{chen2021evaluating}, rely on causal language modelling (CLM)~\cite{feder2021causalm} for pre-training, processing data from left to right while maintaining a transformer decoder structure. Seq2Seq models like CodeT5~\cite{wang2023codet5+} and CommitBART~\cite{liu2022commitbart} employ encoder and decoder mechanisms, with the latter focusing on GitHub commit messages.
The advent of large language models (LLMs) such as ChatGPT~\cite{chatgpt} further stimulated innovation, resulting in specialized coding models like StarCoder~\cite{li2023starcoder}, CodeLlama~\cite{roziere2023code}, and WizardCoder~\cite{luo2023wizardcoder}. These models generally utilize transformer decoders, while CodeT5+~\cite{wang2023codet5+} retains an encoder-decoder setup. These LLMs excel in program repair, code generation, and summarization tasks.
Before the emergence of LLMs, software engineering primarily applied deep learning models in two ways: fine-tuning them~\cite{too2019comparative} in combination with specific tasks or using them as feature extractors without altering their weights. Post-LLM, the trend shifted toward contextual learning~\cite{suryawati2017contextual}, placing these models within broader workflows and integrating domain knowledge for specific scenario applications. To better enable LLMs to solve specific tasks in the future, methods based on efficient fine-tuning of additional weights like LoRA~\cite{hu2021lora} have gained popularity. These methods also include adapter learning~\cite{houlsby2019parameter}, prefix, and prompt learning~\cite{jia2022visual}.
Recent empirical studies~\cite{xiong2023can} on LLMs in the software engineering domain demonstrate how LLMs can reshape methods in the field. 
Tools like MetaGPT~\cite{hong2023metagpt} for automated code generation exhibit the potential for automating software development through sophisticated LLM Agents. Such unit testing efforts also employ iterative prompt engineering to generate test cases and ensure software system quality~\cite{wang2024software}.

\subsection{Federated Learning}

Federated learning, first introduced by Google in 2016~\cite{konecny2016federated}, is a distributed learning approach to machine learning. It enables multiple devices or servers to train data in a distributed fashion without the need to share data samples. This approach significantly boosts data privacy and security since all data remains on the local devices or servers at the data source, eliminating the need for data migration. Federated learning is particularly suitable for processing data that contains sensitive information~\cite{choudhury2019differential}, such as personal health records or personal identification information.

In contrast to traditional centralized machine learning, which involves uploading all local data to a single server, or classical distributed learning methods that assume local data follows the same distribution, federated learning allows multiple parties to collaboratively develop a robust machine learning model without any data sharing. Federated learning begins with a central server initializing a global model. This model is then distributed to various clients, who train it locally with their own data. These clients then send their model updates back to the server, which aggregates these updates to improve the global model. This cycle repeats until the model performs satisfactorily across all clients and converges to a stable state. This way, it effectively tackles the critical challenges of data privacy, data security, data access rights, and handling heterogeneous data.

\noindent
\paragraph{\textbf{Federated Learning Categories}} Federated learning can be divided into three categories based on data application scenarios: horizontal federated learning~\cite{mcmahan2016federated}, vertical federated learning~\cite{yang2019federated}, and federated transfer learning~\cite{pan2009survey}. Horizontal federated learning is applied to data with similar feature spaces but different sample spaces; vertical federated learning is applied to data with the same sample space but different feature spaces; federated transfer learning is applied to data with different sample spaces and different feature spaces. 

Federated Learning also has produced a new form of learning: a fully decentralized learning framework based on blockchain~\cite{li2020blockchain}. Blockchain-based Federated Learning~(BCFL) integrates blockchain technology and federated learning\cite{wang2021blockchain}, aiming to solve issues such as privacy, reliability, and scalability that traditional machine learning frameworks cannot fully accommodate. In BCFL, the central server is replaced by a decentralized blockchain peer-to-peer system, allowing FL to be deployed securely and efficiently. By taking advantage of the unique properties of the blockchain, such as being tamper-proof, auditable, and decentralized, BCFL can guarantee the integrity and safety of the FL process. To enhance the reliability of the global model, BCFL has also introduced an incentive mechanism to reward honest participants and punish dishonest nodes. In the collaborative development environment of open-source AI code models, BCFL is possible to play a significant role. It is entirely decentralized, replacing the central node with a peer-to-peer network~\cite{ramaswamy2005distributed}, ensuring that no superuser can influence the decision-making of the open-source community. Moreover, the blockchain-based FL can effectively reward active participants.

%% file: sections/motivation.tex
\begin{figure}
    \centering
    \includegraphics[width=0.46\textwidth]{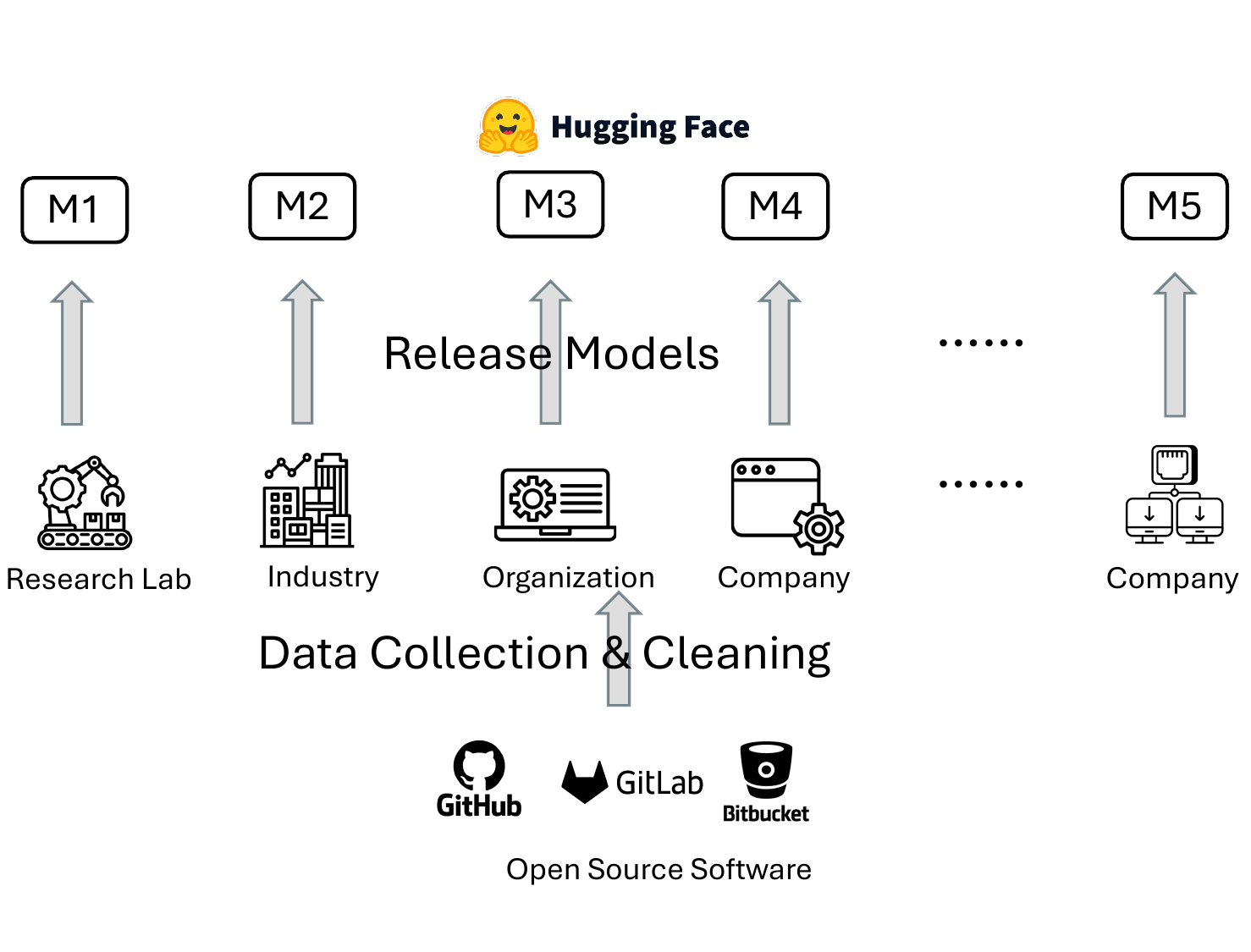}
    \caption{The Current Governance of Open Source Model}
    \label{fig:centralization_oss}
\end{figure}

\begin{figure}
    \centering
\includegraphics[width=0.5\textwidth]{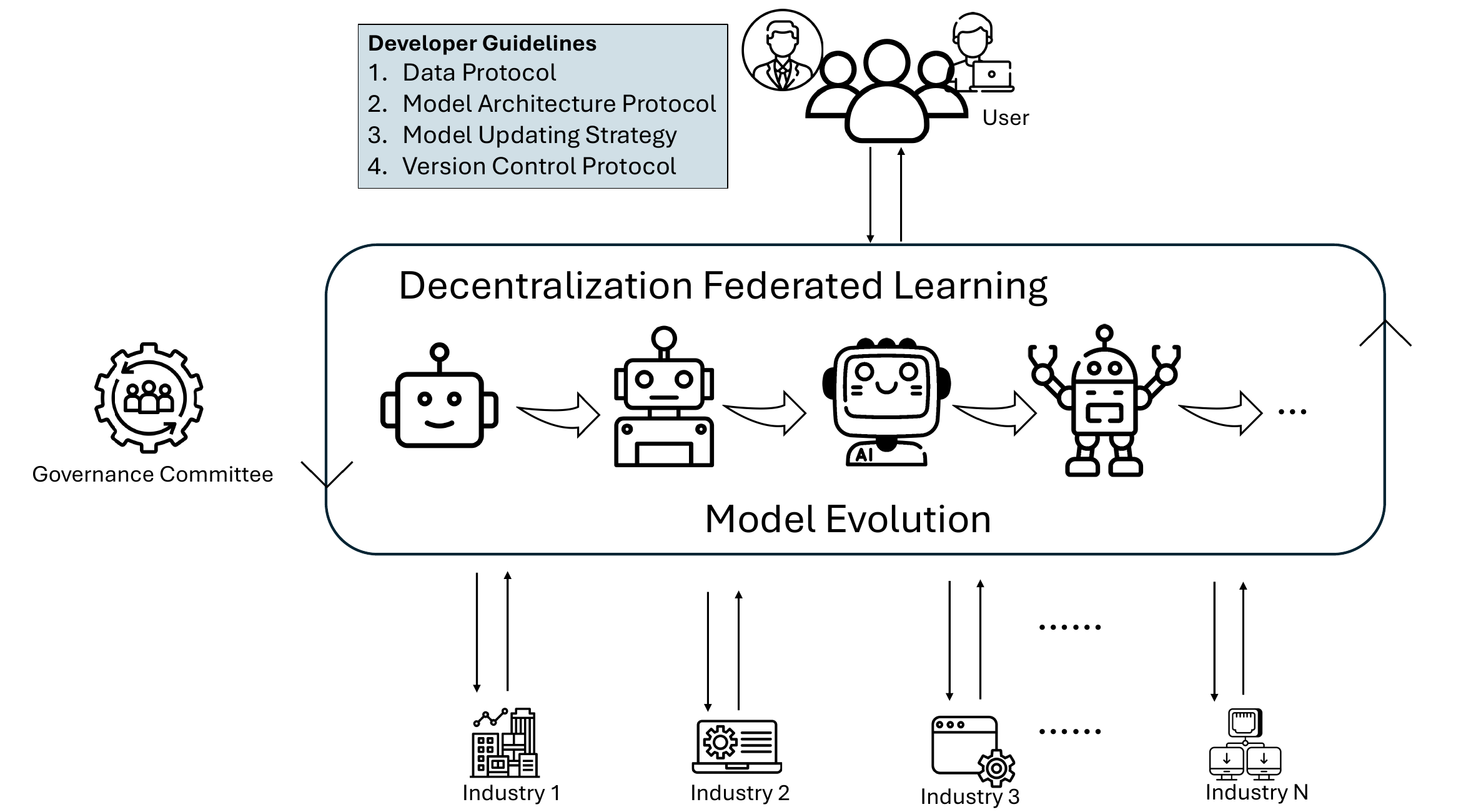}
    \caption{Decentralization Governance of Open Source Model}
    \label{fig:fl_motivation}
\end{figure}
\subsection{Motivation}
Open-source code and collaboration are key drivers of software engineering development. Without this spirit of open collaboration, the IT achievements we see today, even tech giants like Google, would not exist. With the rise and widespread application of artificial intelligence models in software engineering, they have gradually become the basis for research and development works~\cite{amershi2019software}. However, we note that the management process for open-source code models is facing multiple challenges as we mentioned in Section \secref{introduction}.

The traditional method of open-source model dissemination depends on centralized platforms, such as Huggingface~\cite{huggingface}, which allows various teams to share the data and models they have collected and trained as shown in \figref{centralization_oss}. However, this model has not truly realized the possibility of team collaboration in developing the  models. There may be multiple teams training models on the same dataset, leading to duplicate work. Furthermore, the acquisition of training data for open-source code models mainly relies on platforms like GitHub. However, in the era of big data, the data obtained through these methods has limitations in terms of quality and quantity, and cannot fully capture the complexity of coding practices. Further, specific tasks in software engineering, such as bug fixing~\cite{arcuri2008novel} and defect detection~\cite{ngan2011automated} of different language programming, require high-quality data and labels. However, these scarce resources limit our ability to train high-performance AI code models.

After reflecting on the current way of sharing open-source code models, we draw a question: 

\important{\textit{\textbf{Can we co-develop and maintain code models like the development and maintenance methods of open-source software?}}}

The key to AI code models is high-quality data. Unlike the updates in traditional software code, the release of AI code models involves the weight and architecture of the model. \textbf{The biggest challenge} is how to motivate commercial IT organizations to participate in the development of code models and maintain the models while protecting data privacy, which is a key problem we need to solve. Sharing data across organizations may involve sensitive or proprietary code that pertains to the company's core competitiveness.

\subsection{Decentralization Governance of Open Source Model}
Our proposed solution is based on the decentralized federated learning method as shown in \figref{fl_motivation}. Various entities and users collaborate to train a code model using federated learning and this model has the capability to evolve over time. 
To achieve the above target, initially, all participants must reach an agreement on \textbf{the developer guideline}. This guideline covers  \textbf{\textit{data protocol}},  \textbf{\textit{model architecture protocol}},  \textbf{\textit{model updating strategies}}, and  \textbf{\textit{version control protocol}} as explained following:
\begin{enumerate}
    \item Data protocol specifies the data formats for training and inference, including quality standards for data and requirements for test data. 
    \item The model architecture requires participants to conduct local training based on a predefined exemplary model architecture. 
    \item Model updating strategies involve utilizing client weights for model updates, with averaging being a common strategy. 
    \item Version control protocol outlines the appropriate checkpoint saving methods, allowing for saves when model performance reaches a certain threshold on a given dataset. It also defines how to the rules how to control the versions of code models.
\end{enumerate}
\revision{We also need one governance committee to review the pull request that is submitted by the new participants. This committee is responsible to maintain the developer guidelines and manage the community.} 

\revision{In this application scenario of our proposed method, we illustrate this with a bug-fixing model as the case study. Initially, one or several entities establish the data protocol and specify the model's architecture. Subsequently, they choose an existing or devise their own Federated Learning (FL) algorithm. A dedicated committee determines the necessity of launching new iterations when managing versions. They begin by preparing a benchmark dataset. All models poised for release must meet a specific performance criterion and the additional dataset to test the improvement. This performance criterion can include different measurements, such as accuracy and robustness. Every version released must delineate the newly incorporated features. At the outset, collaborative training of the model is undertaken, subsequently making it publicly accessible.} \revision{Later, another team, endowed with more data, discovered this model. Upon evaluation, they acknowledge its efficacy yet note suboptimal performance for specific code data types. They resolve to refine the model, creating a publicly releasable test dataset in the process. Following the established data protocol, this team retrains the model, uploads the refined model and the new test dataset to the repository dedicated to bug-fixing models, and submits a pull request detailing their actions. The governance committee evaluates this submission and decides on its acceptance. Upon approval, the model is updated using the chosen FL algorithm, enhancing its bug-fixing capabilities without compromising existing functionalities.}

This approach addresses key challenges: \textit{privacy protection}, \textit{model generalization}, \textit{collaborative innovation}, and \textit{resource optimization}. By training models locally without exchanging data, federated learning inherently protects the privacy of code data. It also facilitates the merging of local datasets from multiple organizations, ensuring a comprehensive and varied data repository that represents a wide array of programming scenarios. The decentralization of model development promotes extensive collaboration. The strategy of distributing model updates rather than data itself efficiently minimizes unnecessary computations, enabling a more strategic allocation of computational resources.

Therefore, we advocate the advantages of federated learning to build a more inclusive, efficient, and privacy-protected new ecosystem for the development of code models in the field of software engineering. This is not only a technological update but also a strategic adjustment towards a more sustainable and collaborative future direction. We expect the new paradigm to create an environment where competitive advantages come from collective progress rather than isolated development, and technological innovation flourishes while respecting data sovereignty. Under the guidance of federated learning, the development of code models can become more dynamic, responsive, and transparent, adapting better to and reflecting the ever-changing coding practices and needs. Therefore, we look forward to a new age of software engineering—a data-driven, collaborative-centered, and privacy-focused intelligent era.

%% file: sections/experiment.tex
Data is the key driver for the collaborative training and evolution of code models. 
Model performance is strongly related to the data distribution in federated learning~\cite{zhu2021federated}.
Therefore, we designed a series of different data distribution strategies and conducted experiments on 4 code models and 5 code tasks to investigate whether federated learning could significantly reduce or improve model performance. \revision{We designed 3 research questions as shown in \secref{rq} to study the efficiency and effectiveness of FL for AI-based SE tools.}
These experiments aim to verify the feasibility of federated learning in actual software engineering applications and provide solid empirical support for our framework. 

\subsection{Data Distribution}
In this experiment, we designed 6 different data distribution strategies to evaluate the performance and applicability of federated learning:
    \paragraph{\textbf{Benchmark~(Centralization):}} Serving as the control group, this strategy processes all data centrally, training in a single location. It does not involve distributed processing or storage, adhering instead to the traditional centralized machine learning training approach.
    \paragraph{\textbf{Single Client:}} We randomly select a subset as the training data for a single client to explore the difference in model performance when trained with data from only a single client versus using federated learning with multiple clients collaborating. The \textbf{purpose} of this method is to demonstrate that federated learning can enable each participant to improve model performance while ensuring their own data ownership.
    \paragraph{\textbf{Uniform:}} Under this strategy, we ensure data is evenly distributed across all participating nodes, with each client possessing approximately the same number of data points and a relatively balanced distribution of labels.
    \paragraph{\textbf{Label Imbalanced:}} This strategy allows each client to have a similar number of data points but with an uneven distribution of labels within each client. This reflects a common scenario in the real world, where the proportion of data label types varies across different clients, exploring the model's efficiency in handling label-imbalanced data.
    \paragraph{\textbf{Quantity Imbalanced:}} While maintaining consistent label distribution, there are significant differences in the number of samples between clients in this strategy. It simulates the situation in real environments where some clients may have much more data than others, assessing the model's capability to handle such scale discrepancies.
    \paragraph{\textbf{By Repository:}} Data is allocated based on its source repository, ensuring that data from the same user or organization belongs to the same client. This method aims to simulate scenarios of data isolation in reality, such as data processing within a company or organization, to evaluate model performance under such data distribution scenarios.

\revision{We should notice that \textbf{Benchmark~(Centralization)} and \textbf{Single Client} focus on training models in isolation. Conversely, other strategies are specifically related to the nuances of data distribution within the federated learning framework. }

\subsection{Datasets and Baseline Models}
We utilize the CodeXGLUE dataset~\cite{lu2021codexglue}. This dataset is a benchmark collection designed to foster machine learning research for program understanding and generation. 
In our experiment, we choose five tasks: 
\textit{clone detection}, \textit{defect detection}, \textit{code search}, \textit{code-to-text} and \textit{code completion}. Clone detection and defect detection are crucial for maintaining code quality, while code search and code-to-text involve complex interactions between natural language and code. Code completion focuses on improving developer productivity.
These tasks provide an overview of how models perform in real-world programming scenarios, highlighting their abilities in understanding, generating and manipulating code. 

For our experiments, we select the following pre-trained models renowned for their effectiveness in code-related tasks: 1)  CodeBERT~\cite{feng2020codebert} is designed to understand and generate code by leveraging the power of the BERT architecture. It excels in tasks such as code search, code-to-text generation, and more; 2) CodeGPT~\cite{chen2021evaluating} is fine-tuned specifically for programming languages, making it highly capable of code completion and generation tasks; 3) CodeT5 \cite{wang-etal-2021-codet5} is an encoder-decoder model designed to understand and generate programming coded;
4) CodeLlama-7b~\cite{roziere2023code} stands as a large language model specifically tailored for coding tasks. We use it into out framework in order to delve deeper into the potential of federated learning for large model fine-tuning in the code domain. 

\revision{We adopt 4 different, most common federated learning baseline algorithms, FedAvg, FedTrimmedAvg~\cite{yin2018byzantine}, FedMedian and FedProx\cite{li2020federated}. FedAvg aggregates the updates from all participating clients by calculating their average. FedTrimmedAvg helps when client data is noisy or has extreme values. It works by ignoring some of the highest and lowest updates from clients and only using the rest. FedProx modifies the local training on each client by adding a penalty term that measures the difference between the local model and the global model. It helps keep each client's update closer to the overall model, especially when their data is very different. FedMedian uses the middle value of client updates instead of the average. This is useful when client data is very different, as it reduces the impact of very high or very low updates.}

\subsection{Research Questions}
\label{sec:rq}
We have three research questions as shown in the following:
\paragraph{\textbf{RQ1}} \textit{Can participation in federated learning enhance model performance while ensuring data privacy?}  

We compare model performance between single-client training and multi-client federated learning in code-related tasks. This question aims to assess the efficacy of federated learning (FL) in improving model performance over single-client training scenarios for code-related tasks, emphasizing the balance between data privacy and model accuracy.

\paragraph{\textbf{RQ2}} \textit{In code-related tasks, can federated learning achieve performance comparable to centralized training without disclosing data?}

This research question explores the potential of federated learning to approximate the model performance of centralized training setups, thereby evaluating the feasibility of maintaining high model accuracy in the context of data privacy.

\paragraph{\textbf{RQ3}}  \textit{What impact does data heterogeneity have on the performance of federated learning in code-related tasks?}

This question delves into the effects of data heterogeneity on federated learning outcomes, examining how various approaches to data partitioning affect model performance. It aims to explore the challenges and considerations in managing data diversity across clients in federated learning setups. We define 6 data partitioning strategies (e.g., uniform distribution, label-imbalanced, quantity-imbalanced) to study this question.

%% file: sections/results.tex
\subsection{Clone Detection}
\paragraph{Settings} 
We use CodeBERT in this task. We split the dataset into 10 subsets using these distribution strategies: \textit{uniform}, \textit{label-imbalanced}, and \textit{quantity-imbalanced}. 
The training epochs is 2. %

\paragraph{Results} \tabref{code_clone} displaying the results for the Clone Detection task reveals insightful patterns regarding the efficacy of federated learning (FL) settings compared to traditional centralized training approaches across various datasets. Notably, the F1 score under federated learning are closely aligned with those achieved through centralized training when we use FedAVG as the aggregation strategy, indicating that FL is capable of matching the effectiveness of centralized models even when data is distributed across multiple nodes. Furthermore, regardless of the data distribution strategy we use, we find that the models obtained through FL significantly outperform those trained on a single client in terms of F1 scores, demonstrating that participating in federated learning can greatly enhance the performance of the obtained models while ensuring data privacy. 

Additionally, we can compare the effects of different types of data heterogeneity on the results in this task. We find that when the label distribution is uneven, the performance of the models is much lower than that during uniform distribution, indicating certain limitations of federated learning in dealing with label imbalance. However, in scenarios with uneven quantities, the results obtained by the models are even better than those during uniform distribution, suggesting that heterogeneity in the amount of training data does not adversely affect FL and might even enhance the model's robustness.

\begin{table*}[!ht]
\centering
\caption{The F1 score of Clone Detection}
\vspace{-1em}
\label{tab:code_clone}
\begin{tabular}{c|c|c|c|c|c}
\hline

 & \textbf{None-FL}  & \textbf{FedAvg} & \textbf{FedTrimmedAvg} & \textbf{FedMedian} & \textbf{FedProx} \\ \hline
 
Uniform  & - &  \textbf{0.903} & 0.865 & 0.853 & 0.849 \\ %
Label Imbalanced & - & \textbf{0.890} & 0.860 & 0.717 & 0.822 \\ %
Quantity Imbalanced   & - & \textbf{0.912} & 0.857 & 0.820 & 0.854 \\ \hline
Centralization & \textbf{0.941} & - & - & - & -\\ %
Single Client & 0.832 & - & - & - & - \\ \hline
\end{tabular}
\end{table*}

\subsection{Defect Detection}
 \paragraph{Settings} We use Devign dataset which is collected from two large C programming language open-source projects: QEMU and FFmpeg. And we use accuracy as the evaluation metrics. In this task, we use CodeBERT and CodeT5. We split the dataset into 2 parts in order to simulate the realistic scenarios, each part contains one project. 
 We set training epochs to 5.

\paragraph{Results} \tabref{defect_detection} presents the outcomes for the Defect-Detection task. It is evident that under federated learning conditions, both CodeBERT and CodeT5 achieve results that closely rival those of centralized training across this dataset, surpassing the performance seen with training on a single client. 
When we use different strategies to aggregate the parameters, we find that the strategies such as FedTrimmedAvg and FedProx assist in obtaining a more refined model.
\begin{table}[!ht]
\centering
\caption{The accuracy of Defect Detection}
\label{tab:defect_detection}
\scalebox{0.7}{
\Large
\begin{tabular}{c|c|c|c}
\hline
\textbf{} & \textbf{Aggregation Strategy} & \textbf{CodeBERT} & \textbf{CodeT5}\\ \hline
Centralization & - & \textbf{62.08} & \textbf{62.52} \\ %
Single Client & - & 57.54 & 55.53 \\ \hline
\multirow{4}{*}[0ex]{FL(By Repository)} & FedAvg & 61.57 & 58.67 \\ 
 & FedTrimmedAvg & \textbf{62.04} & 59.96\\  
 & FedMedian & 61.53 & 57.65 \\  
 & FedProx & 61.31 & \textbf{61.05} \\  \hline
\end{tabular}}
\end{table}

\subsection{Code Search}
  \paragraph{Settings} In the code search task, we use two models, CodeBERT and CodeT5. Each of these models underwent a triad of training methodologies: centralized, which utilized the complete dataset for training; federated learning (FL), where training was distributed across 10 clients; and single-client training to understand performance under data-constrained conditions. The dataset was split into 10 subsets, with a special emphasis on keeping all repositories from one user within the same subset, to closely mirror practical use cases. We set the training epoch to 2.

\paragraph{Results} In our study on code search as shown in \tabref{code_search}, CodeBERT and CodeT5 are evaluated across centralized, federated learning (FL), and single-client setups. CodeBERT demonstrates an advantage in FL, surpassing its centralized training performance, highlighting its compatibility with distributed data scenarios. Conversely, CodeT5 shows that whatever the aggregation strategies it uses, it performance slightly declined in FL compared to centralized training. Both models experienced a drop in effectiveness when trained on data from a single client, indicating the importance of diverse data sources. 
\begin{table}[!ht]
\centering
\caption{The MRR of Code Search}
\label{tab:code_search}
\Large
\scalebox{0.7}{
\begin{tabular}{c|c|c|c}
\hline
\textbf{}& \textbf{Aggregation Strategy}  & \textbf{CodeBERT} & \textbf{CodeT5}\\ \hline
Centralization & - & 0.2719 & \textbf{0.1951}\\ %
Single Client & -  & 0.2520 & 0.1717 \\ \hline
\multirow{4}{*}[0ex]{FL(By Repository)}& FedAvg &  0.2989 & \textbf{0.1797} \\
 & FedTrimmedAvg &  0.268 & 0.1485 \\ 
 & FedMedian &  \textbf{0.3022} & 0.126 \\ 
 & FedProx &  0.2695 & 0.1478 \\  \hline
\end{tabular}}
\end{table}

\subsection{Code-to-Text}
\paragraph{Settings} In our Code-to-Text experiment, we evaluated the efficacy of CodeBERT and CodeT5 models under 3 distinct training paradigms: centralized training, federated learning (FL) segmented by repository, and training on a dataset from a single client. The federated learning approach aimed to mirror real-world data distribution by dividing the dataset into ten subsets, based on repository names, ensuring that all repositories from a single user were grouped together. We match the federated learning rounds to the centralized learning epochs, setting both to ten. 

\paragraph{Results} \tabref{code_text} shows that centralized training consistently yields the highest bleu-4 score for both models across all languages, underscoring the benefits of extensive data diversity and volume. Federated learning, while designed to maintain data privacy, sees a moderate decline in bleu-4 score, highlighting the challenges of aggregating decentralized learning effectively. Nonetheless, FL outperforms the single-client scenario, indicating that collaborative learning is preferable to isolated data training, reinforcing the notion that data diversity is crucial for model efficacy. Besides, the impact of different aggregation strategies on the outcomes is minimal in this task.

\begin{table*}[]
\centering
\caption{The bleu-4 score of Code-to-Text}
\label{tab:code_text}
\scalebox{0.8}{
\large
\begin{tabular}{c|c|c|c|c|c|c|c|c}
\hline
\textbf{} & \textbf{} & \textbf{Aggregation Strategy} & \textbf{ruby} & \textbf{javascript} & \textbf{go} & \textbf{python} & \textbf{java} & \textbf{php}\\ \hline
\textbf{} & Centralization & - & \textbf{12.16} & \textbf{14.90} & \textbf{18.07} & \textbf{19.06} & \textbf{17.65} & \textbf{25.16} \\ %
\textbf{} & Single Client & - & 7.97 & 9.99 & 14.94 & 15.25 & 13.86 & 20.63 \\  \cline{2-9} 
\textbf{CodeBERT} & {\multirow{4}{*}{FL(By Repository)}}  & FedAvg & 8.90 & 11.54 & 16.90 & \textbf{17.23} & \textbf{16.28} & 23.80 \\ %
\textbf{} & & FedTrimmedAvg & 8.83 & 11.52 & \textbf{16.92} & 17.20 & 16.16 & 23.76 \\ 
\textbf{} &  & FedMedian & \textbf{8.93} & \textbf{11.58} & \textbf{16.92} & 17.21 & 16.15 & \textbf{23.84} \\ 
\textbf{} &  & FedProx & 8.75 & 11.57 & 16.89 & 17.19 & 16.22 & 23.82 \\ \hline
\textbf{} & centralized & - & \textbf{10.75} & \textbf{11.92} & \textbf{14.02} & \textbf{14.80} & \textbf{15.41} & \textbf{23.24} \\ %
\textbf{} & Single Client & - & 9.53 & 9.73 & 12.81 & 11.89 & 12.44 & 17.19 \\ \cline{2-9} 
\textbf{CodeT5} & {\multirow{4}{*}{FL(By Repository)}}  & FedAvg & 9.95 & 9.99 & 13.48 & 12.00 & \textbf{12.85} & 20.56 \\ 
\textbf{} & & FedTrimmedAvg & 9.94 & 9.95 & 13.47 & 11.98 & 12.82 & \textbf{20.57} \\ 
\textbf{} &  & FedMedian & 9.96 & 9.97 & 13.48 & 11.99 & 12.81 & 20.52 \\ 
\textbf{} &  & FedProx & \textbf{9.99} & \textbf{10.00} & \textbf{13.50} & \textbf{12.03} & 12.84 & 20.60 \\ \hline
\end{tabular}}
\end{table*}

\subsection{Code Completion}
\begin{table}[]
\centering
\caption{The accuracy of Code-Completion-Token}
\label{tab:code_completion_token}
\Large
\scalebox{0.765}{
\large
\begin{tabular}{c|c|c|c}
\hline
\textbf{} & \textbf{Aggregation Strategy} & \textbf{CodeBERT} & \textbf{CodeLlama-7b} \\ \hline
Centralization & - & \textbf{76.79}  & \textbf{83.84} \\ %
Single Client & - & 70.22  & 81.59 \\ \hline
 & FedAvg & 73.45  & 83.77 \\ %
 \multirow{4}{*}[2ex]{FL(By Repository)} & FedTrimmedAvg & \textbf{73.79}  & - \\ %
& FedMedian & 73.73  & - \\ %
 & FedProx & 73.63  & - \\ \hline
\end{tabular}}
\end{table}
\paragraph{Settings} In the code completion task, our experimental setup aimed to evaluate the performance of CodeBERT and the large language model, CodeLlama-7b, under various training paradigms: centralized training, federated learning (FL) with dataset division by repository to ensure data privacy, and single-client training to simulate data-constrained environments. We use LoRA to finetune the CodeLlama-7b. We set the training epoch to 5.

\paragraph{Results} \tabref{code_completion_token} shows that the centralized training, which is an ideal scenario with access of the entire datasets, yielded the highest accuracy for both models. It's  encouraging to notice that the slight decrease in accuracy from centralized to FL for both models underscores the trade-offs involved in preserving data privacy through distributed training.  Moreover, the further decrease in accuracy from FL to single client shows the benefits of collaborative learning in FL. Remarkably, CodeLlama-7b outperformed CodeBERT in all tested scenarios. Notably, in the federated learning setup, CodeLlama-7b's performance closely approximated its centralized training accuracy, emphasizing the broad capabilities of Federated Learning in effectively fine-tuning large language models (LLMs).

\subsection{Answers to RQs}
\label{sec:answer}
\subsubsection{RQ1}
 \tabref{code_clone} to \tabref{code_completion_token} demonstrate that, across various tasks and models, multi-client federated learning consistently outperforms single-client training in code-related tasks. Take the Defect-Detection task as an illustrative example, where we divided the dataset into two parts based on project names: one containing defect data from qemu and the other from FFmpeg, a large-scale open-source project renowned for its rich and diverse defect information. \tabref{defect_detection} shows that despite the high quality of data in FFmpeg, federated learning still achieved superior results compared to single-client training solely on the FFmpeg defect dataset.

This observation underscores the advantages of federated learning in harnessing the collective power of multiple data sources while ensuring data privacy. This collaborative approach allows companies to leverage their respective datasets without disclosing sensitive information, thereby fostering a secure and privacy-preserving environment for the development of intelligent software engineering.
\answer{1}{Compared to single client training, federated learning can enhance model performance while ensuring data privacy. The participant can get the benefit from FL.}

\subsubsection{RQ2}

Our experimental results indicate that, while there is indeed a performance gap between federated learning and centralized training in most tasks, especially evident in the Code-to-Text task, federated learning demonstrates considerable promise in specific scenarios. Notably, \tabref{code_completion_token} shows that during the fine-tuning of large models, federated learning achieves result that closely resemble those obtained through centralized training. This suggests that federated learning has significant potential in this area. Furthermore, we observe that in specific code-related tasks, such as code search using CodeBERT, federated learning even surpass the performance of centralized training. This finding implies that federated learning exhibits enhanced robustness in tasks like code search.

While federated learning may not always match the performance of centralized training across all code-related tasks, it offers a viable alternative that balances performance and privacy preservation, particularly in scenarios involving large model fine-tuning or specific tasks like code search.
\answer{2}{Federated learning has shown its potential to approach the performance of centralized training in some scenarios, but there still exists a gap.}

\subsubsection{RQ3}

\tabref{code_clone} presents a detailed overview of the model's performance in various data heterogeneity scenarios. When considering the "Uniform Distribution" strategy, we observe a significant drop in F1 score compared to the Benchmark. On the other hand, the "Label Imbalance" strategy exhibits the lowest F1 score. This underscores the negative impact of label imbalance on model performance. Lastly, the "Quantity Imbalance" strategy maintains a slightly improved F1 score compared to Uniform Distribution in 2/4 aggregate strategies, indicating that the impact of quantity imbalance heterogeneity on clone detection model performance is relatively minor.

In summary, the result reveals that data heterogeneity can significantly influence the performance of models. Therefore, it is crucial to explore techniques and methods that can enhance the model's ability to handle diverse and imbalanced data distributions, thereby improving the overall accuracy and reliability of model.

Besides, as \tabref{defect_detection} and \tabref{code_completion_token} shown, by employing an effective aggregation strategy that mitigates data heterogeneity, FL can attain performance closer to that of centralized training. There is potential for us to identify aggregation strategies that not only achieve performance close to centralized training but could possibly even surpass it.

\answer{3}{Data heterogeneity, particularly imbalances in label distribution, can impact model performance. Exploring various aggregation strategies could be effective in mitigating this influence.}

%% file: sections/challenges.tex
The vision of an open-source code model-sharing platform based on a decentralized federated learning system is to attract more entities to contribute their datasets and collaboratively develop and maintain code models. However, it still faces a series of challenges that need to be addressed. Here, we list some important challenges and possible solutions.

\paragraph{\textbf{Code Privacy Protection}} A core advantage of federated learning is its protection of data privacy through local model training and sharing model weights. Nonetheless, the risk of data leakage still exists, such as threats from member inference attacks~\cite{nasr2019comprehensive}. Unlike some traditional sensitive data—where anonymization can be achieved by masking personal information—code data is extremely sensitive for contributors, even if it is just a variable declaration. One current solution is fully homomorphic encryption~\cite{zhang2021survey}, which allows model training on fully encrypted data and encrypts input data during the inference stage. Although this method is secure, it is computationally expensive. Another potential solution involves designing an intermediary medium to base the training of models and sharing of weights on.

\paragraph{\textbf{Reward Mechanisms}} The development of open-source software often relies on the participation and maintenance of volunteers. However, not all individuals or entities are enthusiastic about participating in open-source projects. It is crucial to design reasonable incentive mechanisms to attract them more effectively to collaborate on open-source models. We can leverage the blockchain technique and create a "contribution token" to reward actively participating individuals and entities. For new users, we can provide initial system participation qualifications through an airdrop. The number of tokens held becomes a symbol of their contribution to the open-source community. This symbol may not directly translate into commercial benefits, but it at least ensures that their contribution is recorded on the blockchain and widely recognized and commemorated. Of course, more complex methods can be adopted, such as introducing DAO (Decentralized Autonomous Organization)~\cite{el2020overview} governance and converting contribution tokens into voting power, making it an influential presence in the governance of the open-source community.

\paragraph{\textbf{Collaborative Interaction Protocols}} When multiple parties collaborate to build and maintain open-source AI models, it is necessary to clarify the rules and protocols for collaborative work. These protocols should cover aspects such as model architecture, data formats, and methods for model updates. New contributors must follow the established rules. The open-source community needs a set of standards to guide the formulation of these protocols.

\paragraph{\textbf{Copyright Issues}} When multiple parties participate together, the ownership of the model may need to be clarified. If there are no clear ownership regulations, this could affect the enthusiasm of participants. To solve this problem, we believe there are two possible approaches: one is to establish copyright ownership rules while formulating collaborative interaction protocols; the other is to share copyrights based on contributions, where the contribution can be measured by the "contribution tokens" mentioned in the reward mechanism.

\paragraph{\textbf{Security issue}} If malicious participants contribute invalid model weights and data, it could seriously damage the open-source ecosystem and contaminate existing models~\cite{yao2024survey}. The reward mechanism implemented based on smart contracts could also contain security vulnerabilities, possibly being exploited by hackers to maliciously acquire a large number of tokens~\cite{8976179}. Moreover, if there are particularly active entities in the open-source community, their continuous increase in token numbers might eventually lead them to a dominant position. It is crucial to design corresponding security mechanisms to prevent these security issues.